\documentclass[conference]{IEEEtran}

\usepackage{tabularx}
\usepackage{booktabs}
\usepackage{multirow}
\usepackage{hyperref}
\usepackage{cite}
\usepackage{amsmath,amssymb,amsfonts}
\usepackage{algorithmic}
\usepackage{graphicx}
\usepackage{textcomp}
\usepackage{xcolor}
\usepackage{pdflscape}
\usepackage{pifont}
\def\BibTeX{{\rm B\kern-.05em{\sc i\kern-.025em b}\kern-.08em
    T\kern-.1667em\lower.7ex\hbox{E}\kern-.125emX}}
\begin{document}

\title{Interoperability Effects: Extending DeFi Lending Risk Models to Multi-Chain Environments}

\author{\IEEEauthorblockN{Hasret Ozan Sevim}
\IEEEauthorblockA{
\textit{University of Camerino (Camerino, Italy) \& Catholic University of Sacred Heart (Milano, Italy)}\\
\texttt{hasretozan.sevim@unicam.it}
} \\
}

\maketitle

\begin{abstract}
On-chain lending has expanded across multiple distributed ledgers as DeFi becomes increasingly multi-chain. This environment introduces novel technical and financial mechanisms, particularly cross-blockchain communication and asset transfer protocols, yet cross-chain elements remain understudied in lending protocol risk management. To address this gap, we applied panel regression fixed effects and OLS models to empirically analyze the impact of cross-blockchain interoperability solutions, using TVL and total revenue as performance proxies from October 2022 to January 2025. Our data set covers 15 decentralized lending protocols and 53 cross-chain bridges across 9 EVM-compatible blockchains, categorized as Ethereum, alternative layer-1s, and Ethereum layer-2 networks. The results reveal that cross-chain activity impacts on protocol performance. Bridge volume emerges as a significant explanatory variable. It shows a significant association with TVL and revenue across different categories, though the direction of this effect varies heterogeneously. Increased bridge integrations are generally associated with decreased TVL and protocol revenue across categories, potentially indicating liquidity escapes from these lending ecosystems. Liquidations produce heterogeneous effects across categories. The panel data reveal that new network launches are weakly associated with TVL and revenue, while bridge hacks show no significant relationship. The R-squared values confirm meaningful explanatory power. We further show that Ethereum attracts large depositors, while layer-2s skew toward retail participation. We conclude that effective DeFi risk models should incorporate cross-chain metrics and adopt a layer-aware approach to accurately reflect the evolving multi-chain landscape.
\end{abstract}

\begin{IEEEkeywords}
Decentralized Finance, Blockchain, Lending, Risk Management.
\end{IEEEkeywords}

\section{Introduction}

Decentralized finance (DeFi or 'on-chain finance') is a growing financial system built on distributed ledger and smart contract technologies. DeFi promises a more transparent, interoperable, secure and peer-to-peer transacting financial environment, with less inclusion of classical financial intermediaries \cite{Schaer2021}. One of the leading DeFi services is automated lending markets (ALMs), also known as decentralized lending protocols. At the core, these protocols are smart contracts programmed for asset depositing, borrowing and liquidation (in case of insolvency) with similar abilities of financial intermediaries like a bank. Currently, DeFi services are available on multiple blockchains where ALMs operate. This creates a difference in terms of liquidity, performance, speed, and cost between lending services on different networks. More importantly, this fragmentation in DeFi has brought cross-chain 'interoperability' products ('bridge' protocols) \cite{Harris2023}. In January 2026, cross-chain asset transfer protocols have transferred more than 160 billion USD worth of crypto-assets on different blockchains and create a new dynamic for DeFi environments \cite{DeFiLlama}.

\begin{figure}[h]
    \centering
    \caption{illustrates the evolved multi-blockchain on-chain finance ecosystem.}
    \vspace{0.05cm}
    \includegraphics[width=0.48\textwidth]{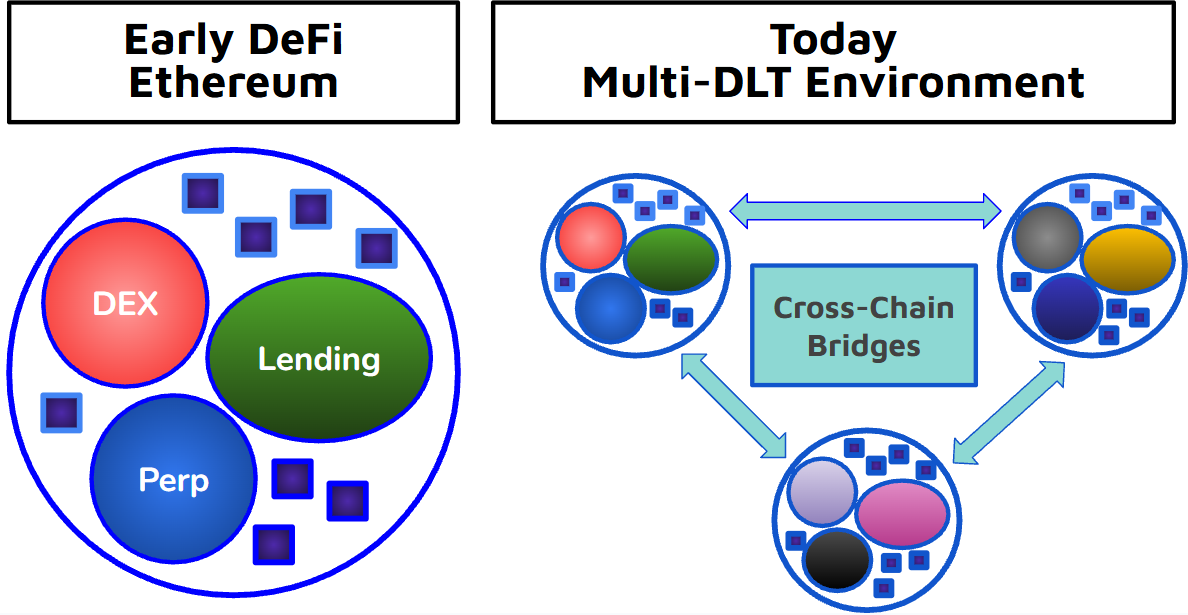}
    \label{fig:diagram2}

\vspace{0.3em}
\raggedright
\scriptsize
The interconnectedness created by cross-chain bridges allows cascades across lending ecosystems on separate blockchains. A major shock, such as a bridge hack draining assets, a new network launch shifting liquidity, or a sudden spike in bridge volume, can trigger the depegging of bridged assets and create bad debt in one network's lending protocols. Because these protocols on different chains are linked via shared collateral and liquidity through bridges, this insolvency rapidly transmits to the others, causing a wave of cross-chain liquidations. This contagion results in a synchronized and structural break, sometimes even a simultaneous crash in TVL and protocol revenue across all connected networks, demonstrating that their financial stability is not isolated but interdependent.
\end{figure}

DeFi faces significant risks, including regulatory concerns, scalability issues, liquidity risk, oracle risk, credit risk, and smart contract vulnerabilities. The liquidity flows through cross-chain bridge contracts adds another risk layer that can be named as 'bridge risk'. All risks also manifest in DeFi lending markets, with the potential to cause considerable losses to participants and protocols. The total value locked in DeFi reached about \$120 billion, with the significant contributions of lending platforms such as Aave and MakerDAO \cite{DeFiLlama}. Analyzing the health and stability of lending protocols in response to these risks is crucial to understanding the automated risk management offered by different protocols, so that investors and policymakers can make informed decisions. In particular, we find that the protocol and market data-driven research of lending risk management with multi-blockchain approach are limited in the literature. There are studies in the literature with the findings indicating that liquidation events can impact on the TVL of lending protocols \cite{Luo2025}, and analyzing the impact of liquidation over the TVL and revenue on different protocols and blockchains \cite{Iftikhar2025}. However, we find that previous studies have not compared the performance and automated liquidity risk management of different lending environments with the elements of cross-chain asset transfers although the bridge protocols have become essential for liquidity management. Consequently, we are currently in an underexplored and poorly optimized cross-ledger (or 'cross-chain') financial environment. We aim to integrate cross-chain bridge activity into DeFi risk models to better understand its impact on protocol stability and performance.

This work compares multiple blockchain (multi-chain) types, capturing the nuances of liquidity management, protocol performance and cross-chain engagements across both layer-1 and layer-2 networks, using data covering the period between 17 October 2022 and 1 January 2025. We applied a panel data analysis with fixed effects and OLS regression models to discover fresh insights into cross-chain performance and the risk management of liquidity. We used TVL and total revenue (TR) as proxies for protocol stability, regressed to cross-chain bridge-related and liquidation-related variables. The research provides empirical insights of critical factors indicating influence on the growth and resilience of decentralized lending platforms.

\begin{table}[t]
\centering
\caption{Blockchain Network Taxonomy for the Analysis}
\label{tab1}

\begin{tabularx}{\linewidth}{l l c l l}
\toprule
\textbf{Category} & \textbf{Blockchain} & \textbf{Launch} & \textbf{Protocols} & \textbf{Bridges} \\
\midrule

L1 &  Ethereum & 30-07-2015 & 9 & 53 \\

\midrule
\multirow{6}{*}{L2}
 &  Polygon & 01-05-2020 & 4 & 34 \\
 & Arbitrum One & 31-08-2021 & 4 & 39 \\
 &  Optimism & 16-12-2021 & 3 & 36 \\
 &  Gnosis & 08-12-2022 & 2 & 7 \\
 &  zkSync Era & 24-03-2023 & 1 & 18 \\
 &  Base & 09-09-2023 & 2 & 34 \\

\midrule
\multirow{2}{*}{AltL1}
 & BNB & 01-09-2020 & 4 & 32 \\
 &  Avalanche & 21-09-2021 & 5 & 26 \\
\bottomrule
\end{tabularx}
\raggedright
\scriptsize
This table shows the taxonomy of the blockchain networks as of 1st January 2025, with the number of existing lending protocols (\textit{Protocols}) and bridge protocols (\textit{Bridges}), dates of the network launches (\textit{Launch} in DD-MM-YYYY format), the \textit{Category} of the network. Source: DeFiLlama (Last Access: January 1, 2025).

\end{table}

\begin{figure}[h]
    \centering
    \caption{Cross-Blockchain Bridge Volume by Network Groups}
    \vspace{0.05cm}
    \includegraphics[width=0.48\textwidth]{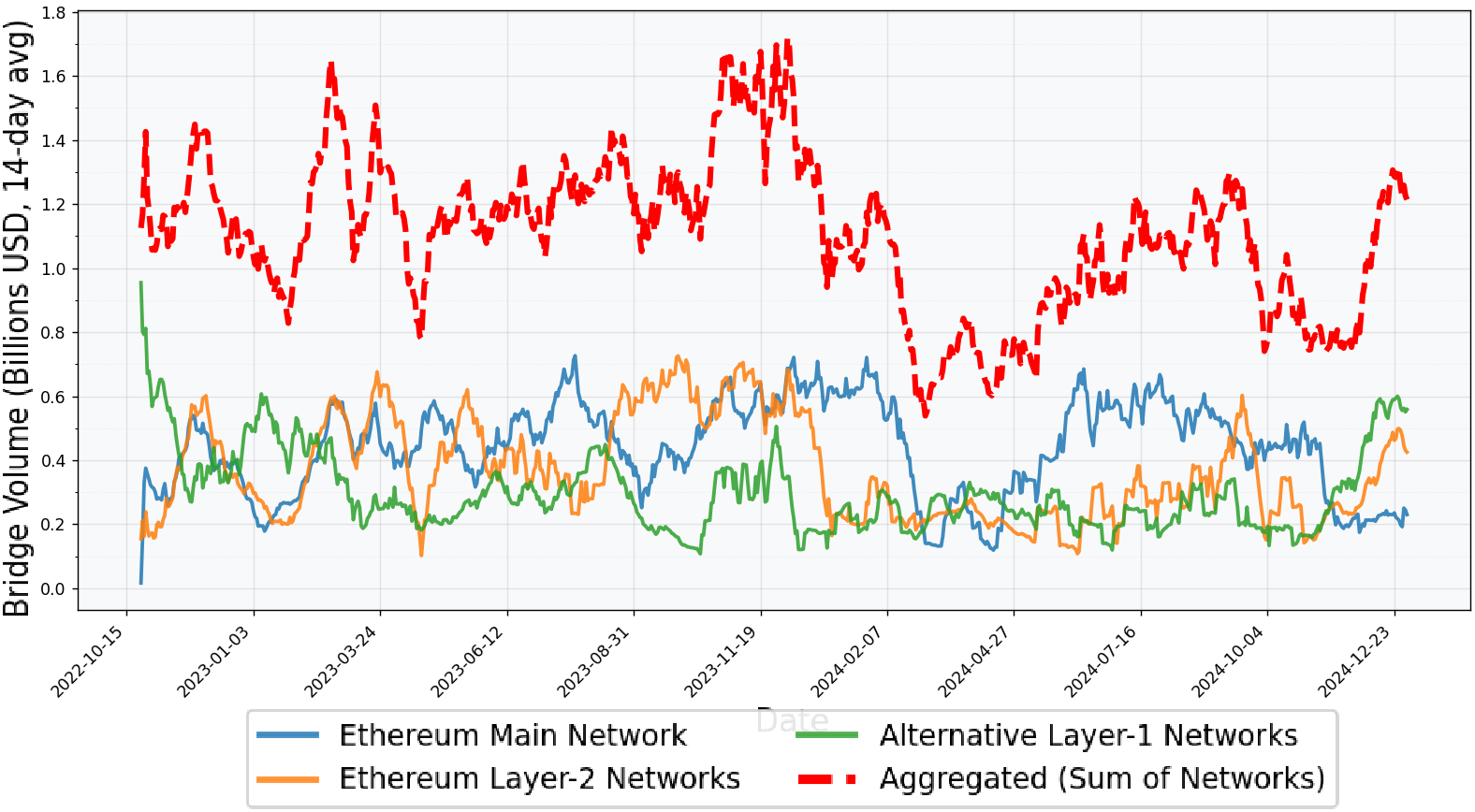}
    \vspace{-0.15cm}
    \label{fig:bridgevolume_plot}
\end{figure}

\begin{figure}[h]
    \centering
    \caption{Multi-Blockchain TVL in the Lending Protocols by Network Groups}
    \vspace{0.05cm}
    \includegraphics[width=0.48\textwidth]{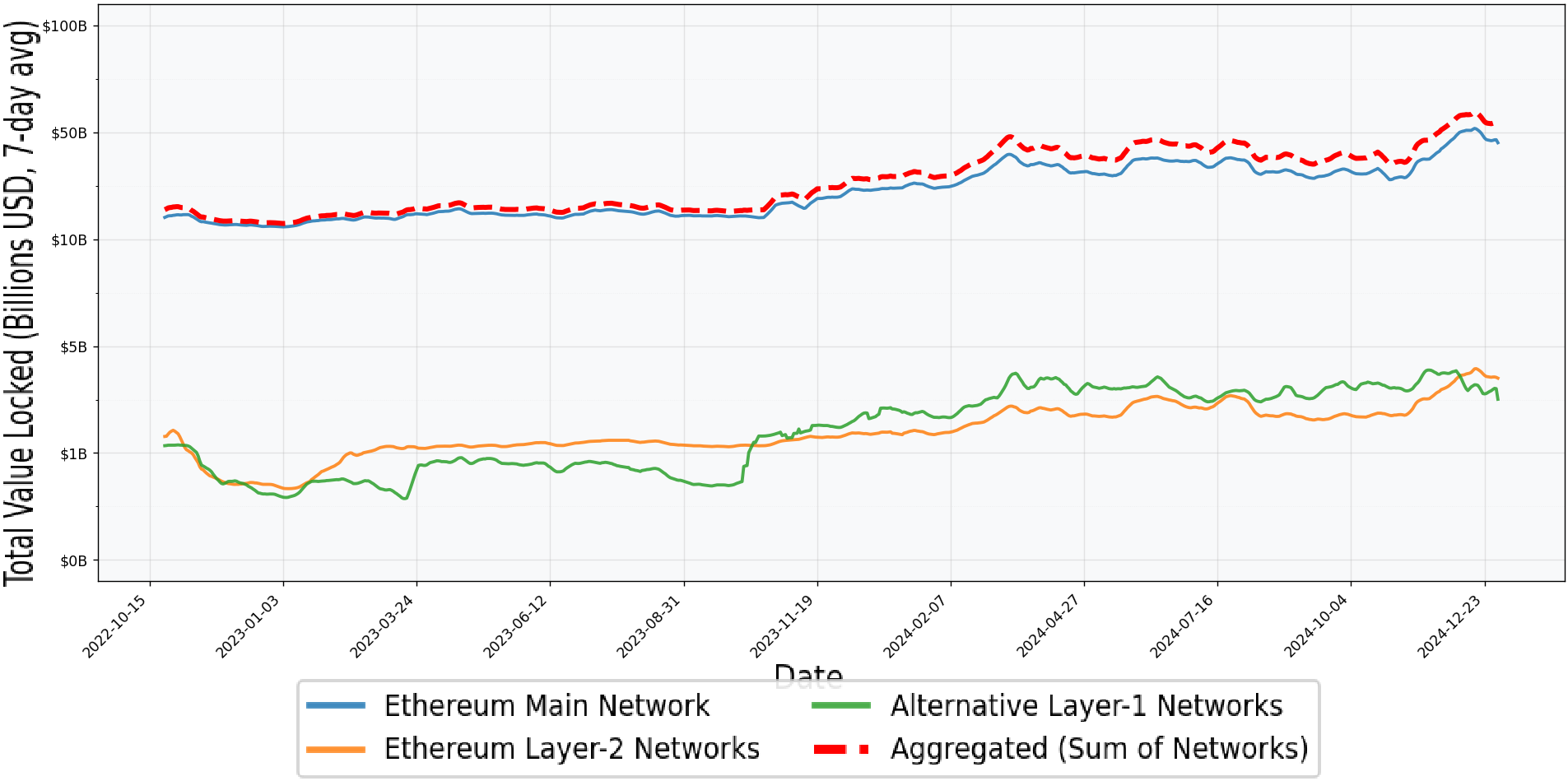}
    \vspace{-0.15cm}
    \label{fig:tvl_plot}
\end{figure}

\section{Background}

Risk management in lending protocols relies on automated mechanisms like over-collateralisation ratios, governance frameworks, and liquidation triggers to protect participants and maintain stability. However, these measures can be insufficient during periods of high volatility, market congestion, liquidation spirals, sequencer and bridge outage, and systemic risk. High gas prices can also reduce their effectiveness in maintaining stability \cite{LeharParlour2022} \cite{Ante2024} \cite{Bertomeu2024}. The existence of different lending protocols on different networks and liquidity rails including cross-chain asset transfers facilitate more stability in lending and liquidation operations, such as collateralization modifications. Our study focuses on investigating the risk parameters that exert a direct impact via cross-chain asset transfers, liquidations, and control variables.

\cite{Qin2021}, \cite{Warmuz2022}, \cite{LeharParlour2022}, \cite{Cohen2023} and \cite{Kao2020} investigated liquidation mechanisms, liquidation dynamics, and fragility in DeFi lending markets. \cite{Sun2023} proposes an estimation model to measure liquidity risk. \cite{Saengchote2022} investigates the concentration of deposits and loans among users to assess systemic risk within the protocol with an estimation model.

Previous studies, with a particular focus on the relationship between liquidations and TVL, highlight the negative
impact of liquidations on the TVL of the protocols. \cite{Luo2025}, \cite{Sun2023}, \cite{Mottaghi2025}, \cite{Grigorova2024}, and \cite{Metelski2022} contributed with various findings on the TVL, total revenue, liquidation and their relationships. \cite{Sinyugin2023} contributed descriptive insights, including mass liquidations can sometimes signal efficient risk management. \cite{Luo2025} investigated the impact of ETH volatility on TVL of the protocol via liquidations. They performed an empirical and sensitivity analysis using the decline in ETH as a price shock to TVL, while controlling different factors, such as the gas price, the liquidation threshold, and VIX. Their findings showed that TVL is highly unstable during market downturns due to cascading liquidation events. They used different performance metrics including TVL and total revenue (TR) and confirmed the negative impact of liquidations on TVL, but found insignificant results for the impact on TR.

\cite{Iftikhar2025} contributed a cross-protocol risk management model  to the growing literature that focuses on the efficiency of risk management system in upgraded versions of lending protocols by investigating protocol’s advanced features, transaction costs, volatility, and sentiment in crypto and traditional markets. The study evaluated different versions of Aave and Compound lending protocols across Ethereum and Arbitrum layer-2 network.

\cite{Mottaghi2025} shows that volatility patterns of DeFi are significantly influenced by the structural distinction between Layer 1 and Layer 2 blockchains. The outcomes of the study show the importance of network selection, important strategic risk management, and adaptive liquidity techniques to improve stability in distributed financial ecosystems.

Despite the growing importance and volume of cross-chain activities, existing studies didn't involve the cross-chain elements in the models. Our study adds to the growing literature by analyzing the relationship between the stability of decentralized lending and cross-chain elements.

This study addresses the following research question: Do the cross-chain activities (cross-chain asset transfer volume, bridge hacks, bridge integrations, and new layer-2 launches) have a statistically significant impact on the total value locked and revenue of on-chain lending environments?

\section{Data}

We collect protocol data from The Graph\cite{TheGraph}, which is a platform for querying blockchain data by ensuring that all data were sourced directly from transactions recorded on-chain. The use of subgraphs allowed us to focus on specific metrics of the lending protocols, such as daily revenue, TVL, liquidation volume, withdrawal volume, borrowing amount, deposit amount and user counts. This approach ensured a comprehensive and consistent dataset, particularly for protocols operating on multiple networks. Average gas price data of the blockchain networks is collected from the blockchain explorers: EtherScan \cite{Etherscan}, ArbiScan \cite{Arbiscan}, BaseScan \cite{Basescan}, PolygonScan \cite{PolygonScan}, OP Mainnet Explorer \cite{OPMainnetExplorer}, Gnosis Chain (xDAI) Blockchain Explorer \cite{GnosisExplorer}, ZKsync Era Block Explorer \cite{ZkSyncExplorer}, Snowtrace \cite{SnowTrace}, BscScan \cite{BscScan}. ETH price data is collected from EtherScan \cite{Etherscan}. Bridge volume and annual percentage yield (APY) data are collected from DefiLlama \cite{DeFiLlama} while the DeFiLlama data was publicly accessible. The data of the 'Crypto Fear Index \& Greed Index’ is collected from the API of Alternative \cite{AlternativeFearGreed}. Bridge volume, liquidation volume, withdraw volume, revenue, TVL, ETH price and average gas price values are logarithmically transformed. The code and data used during the research can be shared upon request.

\begin{figure}[h]
    \centering
    \caption{Active User Count in Lending Protocols by Network Groups}
    \vspace{0.05cm}
    \includegraphics[width=0.48\textwidth]{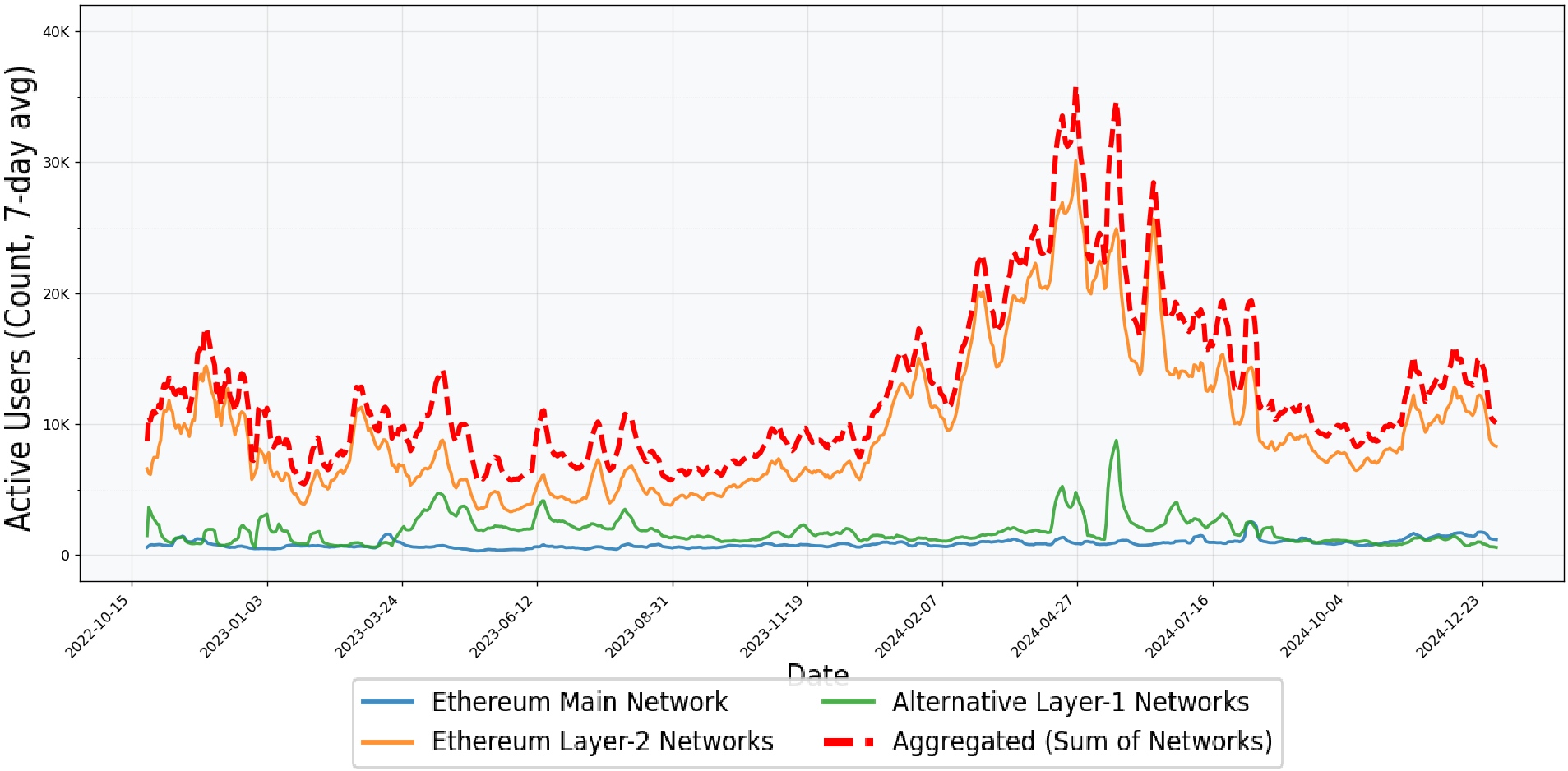}
    \vspace{-0.15cm}
    \label{fig:users_plot}
\end{figure}

The time range of total asset bridge volume data is from 17-10-2022 to 01-01-2025 with 807 observations in daily time intervals. The data of AAVE, MakerDAO, Compound, Spark, Zerolend, Abracadabra Money, Liquity, Banker Joe, BENQI, Venus Protocol, and Sonne Finance are collected from various networks. The list of lending protocols and bridge protocols is presented in Table \ref{tab1}. Dummy event variables are used to observe the structural impact of the cross-chain events. The list of events (bridge hacks, bridge integrations and 'mainnet' network launches) is presented in Table \ref{tab8}. The event dates are collected from the official announcements and declarations of the related protocols and networks.

We classify blockchain networks into three categories: Ethereum mainnet (L1), other layer-1 networks (AltL1), and Ethereum layer-2 networks (L2). This classification reflects differences in architecture, settlement finality, security inheritance, and execution environments that influence protocol composition and financial dynamics \cite{Rossi2023} \cite{Gangwal2022}. There are many studies, indicating different DeFi alignments between Ethereum and rollups \cite{Gogol20252} \cite{Rossi2023}. We investigate the diverse impact of different architectures in lending environments. Estimating alternative chains on its own leads generates distributed and path-dependent effects, sparse events, and few associated DeFi protocols, resulting in fragmented and economically weak insights. Grouping blockchains by shared settlement and performance architecture preserves statistical power while retaining economically meaningful heterogeneity. Our objective is not to compare individual chains or protocols, but to identify how architectural design choices systematically condition exposure to cross-chain risk dynamics. By aggregating networks within each category, our results should be interpreted as architecture-level effects rather than chain-specific performance outcomes.

\begin{figure}[h]
    \centering
    \caption{Deposit Volume in the Lending Protocols by Network Groups}
    \vspace{0.05cm}
    \includegraphics[width=0.48\textwidth]{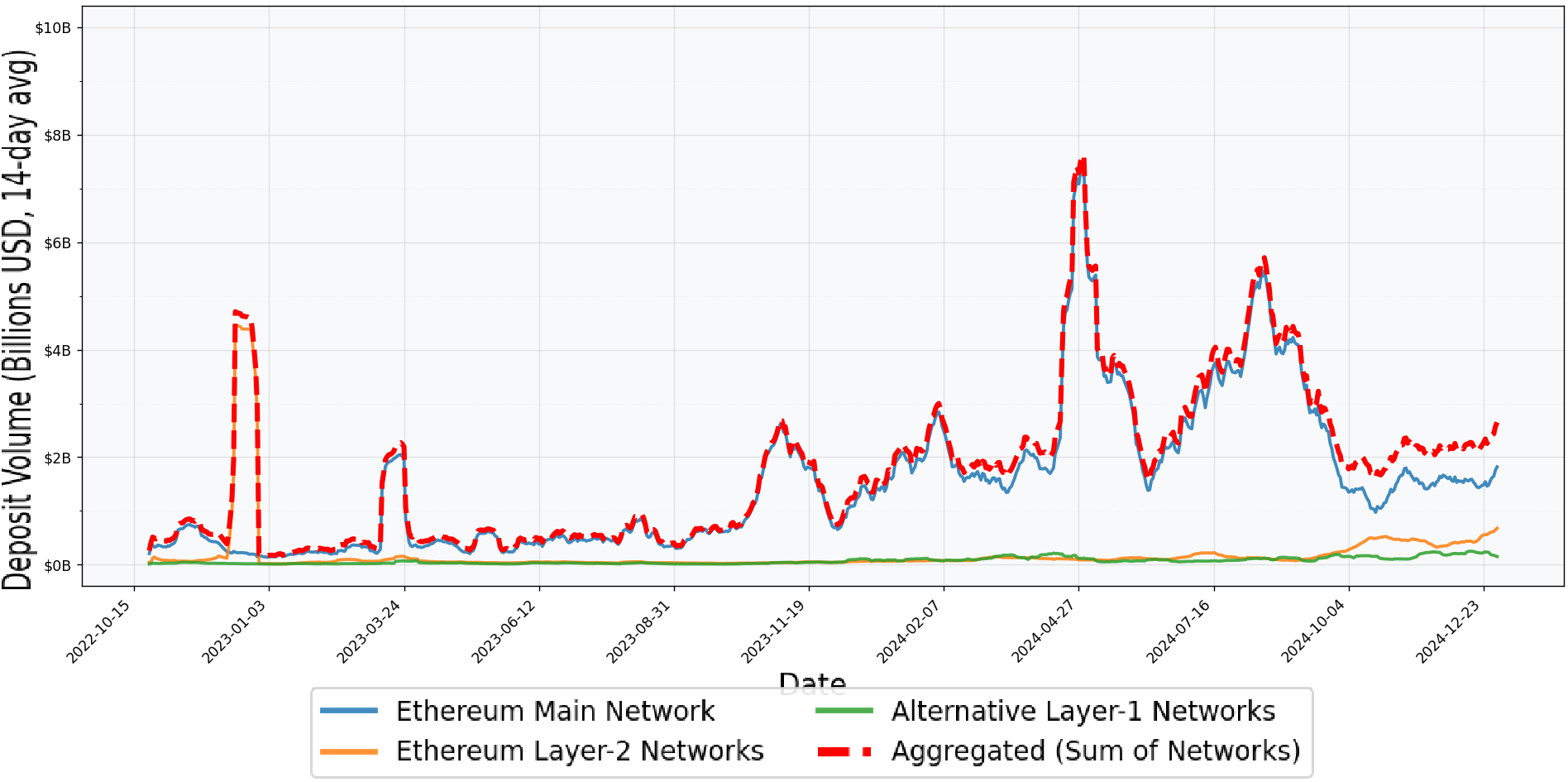}
    \vspace{-0.15cm}
    \label{fig:deposit_plot}
\end{figure}

This study focuses exclusively on Ethereum Virtual Machine (EVM) compatible blockchains for both technical and analytical reasons. From a technical standpoint, EVM chains share a unified smart contract execution environment, meaning lending protocol architectures, liquidity accounting mechanisms, and bridge integration standards are structurally comparable across chains. This enables meaningful cross-chain performance comparisons that would otherwise be confounded by architectural heterogeneity. Secondly, the EVM ecosystem constitutes the dominant share of total DeFi TVL and cross-chain bridge activity, representing the most empirically relevant environment for studying interoperability risks in lending \cite{DeFiLlama}. Thirdly, restricting the sample to EVM chains ensures data consistency: on-chain metrics including TVL, protocol revenue, and liquidation volumes are collected through standardized subgraph queries and indexing infrastructures. The three analytical categories reflect meaningful architectural and economic distinctions within this ecosystem, as established in prior blockchain infrastructure literature, while remaining technically homogeneous enough to support rigorous comparative analysis.

\section{Methodology and Model}

Regression analysis is a common method to determine the relationship between dependent and independent variables in DeFi markets. Studies that have used regression analysis to assess various aspects in DeFi markets include \cite{Gogol2024}, \cite{Moallemi2024}, \cite{Hansson2022} and \cite{Bertucci2024}. In this study, we employ a panel data regression with fixed effects and OLS regression models to discover fresh insights into cross-chain performance and risk management of liquidity. The regression models provide parametric estimates and significance testing \cite{Wooldridge2010}. We used TVL and total revenue (TR) as proxies for protocol stability, regressed to cross-chain bridge-related and liquidation-related variables. The analysis provides empirical findings on cross-chain activities that indicate influence on the growth and resilience of decentralized lending platforms.

We construct a \textit{Credit Expansion Ratio} (CER) defined as 
\[
CER_{n,t} = \log\!\left( \frac{\textit{dailyBorrowUSD}_{n,t}}{\textit{dailyDepositUSD}_{n,t}} \right).
\]
aggregated at the network–type level. This measure captures the relative intensity of borrowing flows compared to deposit flows within a given network-group and day. We follow prior descriptive uses of daily borrow and deposit flow metrics in DeFi research and policy analysis \cite{HeimbachHuang2024}, where such flow-based indicators are interpreted as signals of short-run credit expansion and potential procyclicality rather than as balance-sheet stock measures such as loan-to-value (LTV) ratios.

\begin{align*}
    Perf_{i,t} &= \alpha_{i} + \beta_1 \text{BridgeVolume}_{i,t} \\
    & + \beta_2 \text{BridgeVolume}_{i,t} \text{L2}_{i} + \beta_3 \text{BridgeVolume}_{i,t} \text{AltL1}_{i}  \\
    & + \beta_4 \text{Liquidation}_{i,t} + \gamma_1 \text{CER}_{i,t} + \gamma_2 \text{Withdraw}_{i,t} \\
    & + \gamma_3 \text{volETH}_{i,t} + \gamma_4 \text{GasPrice}_{i,t}  + \gamma_5 \text{ETH\_APY}_{i,t} \\
    & + \gamma_6 \text{Stablecoin\_APY}_{i,t} + \gamma_7 \text{FGI}_{t} + \gamma_8 \text{BridgeHack}_{t} \\
    & + \gamma_9 \text{Integrations}_{t} + \gamma_{10} \text{Mainnet}_{t} + \epsilon_{i,t}
\end{align*}

The baseline regression model is introduced in the equation above. The dependent variable, \(Perf_{t,i}\), measures the performance
of protocols on network-type \( i \) at time \( t \), using either \(TVL_{t,i}\) or \(Revenue_{t,i}\) data after logarithmic transformation. Our main explanatory variables are the logged bridge volume, \(BridgeVolume_{t,i}\), and the logged liquidation volume, \(Liquidation_{t,i}\). The coefficient for \(BridgeVolume_{t,i}\) represents the impact of the bridge volume on Ethereum (the base category), while the coefficients for the interaction terms \(BridgeVolume_{t,i}L2_{i}\) and \(BridgeVolume_{t,i}AltL1_{i}\) represent the additional effect of the bridge volume on L2 and AltL1 compared to Ethereum L1. We include several control variables to account for potential confounding factors. The variable \(CER_{t,i}\) captures short-run credit expansion of the lending protocols on network-type \( i \) at time \( t \). Three dummy variables are incorporated to control potential structural break and discrete exogenous shocks affecting TVL and revenue through cross-chain liquidity channels. So they utilize to isolate the specific effects of the main explanatory variable \(BridgeVolume_{t,i}\). \(Integrations_{t}\) represents cross-blockchain bridge integrations at time \( t \), \(BridgeHacks_{t}\) represents hack of bridge protocols to control major negative external shocks at time \( t \), and \(Mainnet_{t}\) represents new blockchain network launch events at time \( t \). Market volatility is represented by \(VolETH_{t,i}\) for the benchmark collateral asset ETH with 7-day rolling standard deviation of the log-returns. \(GasPrice_{t,i}\) controls the network activity, \(ETH\_APY_{t,i}\) controls the annual percentage yield of
ETH, and \(Stablecoin\_APY_{t,i}\) controls the annual percentage yield of
stablecoins. Protocol-level activity is measured by \(Withdraw_{t,i}\) as the logged total withdrawal volume from the lending protocols on network-type \( i \) at time \( t \). \(FGI{t}\) is incorporated to control for broader market sentiment as a proxy for investor sentiment in
the crypto-asset markets.

To capture heterogeneity of bridge volume across network groups, we introduced two interaction terms. \(BridgeVolume_{t,i} \text{L2}_{i}\) corresponds to the heterogeneous effect of bridge volume on the L2s while \(BridgeVolume_{t,i} \text{AltL1}_{i}\) corresponds to AltL1s.

\(GasPrice_{t,i}\) is a TVL-weighted index of gas cost pressure by aggregating network-level log changes in gas price across networks at time \( t \) on network-type \( i \). The weight of each network is determined by its proportion of the total value locked (TVL) in all networks. The main reason for adopting network-level log changes in gas price calculation is to minimize the impact of different network gas standards across multiple networks.

    \begin{equation}
    \mathrm{GasPrice}_{i,t}
= \sum_{n \in i} \left( \frac{\mathrm{TVL}_{n,t}}{\sum_{m \in i} \mathrm{TVL}_{m,t}} \right)
\Delta \log(\mathrm{GasPrice}_{n,t}) 
    \end{equation}

Similarly, \(ETH\_APY_{t,i}\) is a weighted average of the APY of ETH on protocols at time \( t \) on network-type \( i \). The weight of the ETH APY is calculated according to the TVL in the lending protocols on each network. \(Stablecoin\_APY_{t,i}\) is a weighted average of stablecoin APY on protocols at time \( t \) on network-type \( i \). For each network, the protocol‑level APYs are weighted by their protocol TVL, and the resulting network‑level APY further reflects the TVL‑weighted average of the major stablecoins (USDC, USDT, and DAI).

    \begin{equation}
\mathrm{APY}_{i,t}
=
\sum_{n \in i}
\left(
\frac{\mathrm{TVL}_{n,t}}{\sum_{m \in i} \mathrm{TVL}_{m,t}}
\right)
\mathrm{APY}_{n,t}
    \end{equation}

\section{Results}

Table \ref{tab:summary_statistics} demonstrates the summary statistics of the raw aggregated data. VIF values are presented in Table \ref{tab:VIF}, which indicates no multicollinearity. The variables do not show a mechanical correlation ($>$0.90) in a correlation matrix. Figure \ref{fig:bridgevolume_plot} shows the cross-chain asset transfer volume through bridge volumes over time by network groups. Figure \ref{fig:tvl_plot} shows the TVL in the lending protocols of network groups. The Ethereum network is leading in the TVL metric by far. Figure \ref{fig:deposit_plot} demonstrates the deposit volume in the lending protocols of network groups. Figure \ref{fig:users_plot} demonstrates the user count in the lending protocols of network groups. The lending protocols on the layer-2 networks are used by more users, while the Ethereum network is preferred for big deposits by value.

The results of panel data regression with fixed effects are presented in Table \ref{tab:fixed_effects_regression_results}. The pooled OLS regression results are presented in Table \ref{tab:pooled_ols_regression_results}. The results of the category-specific OLS regression are introduced in Table \ref{tab:network_specific_ols_regression_results_tvl}.

\begin{table}
\centering
\caption{Summary Statistics}
\label{tab:summary_statistics}
\begin{tabularx}{\linewidth}{@{}Xccccccc@{}}
\toprule
\textbf{Metric} & \textbf{Mean} & \textbf{Std. Dev.} & \textbf{Min} & \textbf{Max} \\
\midrule
TVL (\$bn)          & 31.56 & 11.78 &	15.70 & 63.02 \\
Revenue (\$mn)      & 1.43 & 1.08 & 0.23 & 5.77 \\
BridgeVolume (\$mn) & 224.35 & 154.43 &	17.83 & 1334.51 \\
Liquidation (\$mn)  & 2.49 & 19.80 & 00.00 & 459.02 \\
Withdraw (\$bn)     & 1.77 & 2.35 & 0.07	& 39.20 \\
CER	                & 0.37 &	0.29 &	0.02 &	3.20 \\
ETH Return (\$)	    & 0.00	& 0.03	& -0.19	& 0.18 \\
ETH APY	            & 0.95 &	0.18 &	0.48	& 3.33 \\
Stablecoin APY      &	5.13 &	2.95 &	1.27	& 15.78 \\
Gas Price           &	0.00   &	0.27 &	-1.73 &	1.74 \\
FGI	                & 55.97 &	17.30 &	17 &	94 \\
\midrule
\textbf{Dummy} & \textbf{Day Count} \\
Bridge Integration Dates	& 14 \\
Network Launch Dates  & 2 \\
Bridge Hack Dates	& 5 \\
\bottomrule
\end{tabularx}
\raggedright
The table includes the summary statistics of the aggregated raw data from 17 October 2022 to 1 January 2025, with outliers removed. Some revenue values had extreme outliers queried from The Graph. These extreme values in the protocol-side revenue (37 values in total) are removed from the calculation of the total revenue. The number of daily observations is 807. There is no missing day.

\end{table}

\begin{table}[htbp]
  \centering
  \caption{Panel Data Regression Results}
  \label{tab:fixed_effects_regression_results}
  \begin{tabular}{lcccc}
    \toprule
    & Model 1 (TVL) & Model 2 (TR) \\
    \midrule
    BridgeVolume            & -0.0606$^{**}$ (0.0207)   & 0.1381$^{***}$ (0.0328)\\
    BridgeVolume:L2         & 0.1498$^{***}$ (0.0251)   & -0.1346$^{***}$ (0.0397) \\
    BridgeVolume:AltL1      & -0.0088 (0.0254)          & -0.2728$^{***}$ (0.0402) \\
    Liquidation              & 0.0006  (0.0023)          & 0.0172$^{***}$ (0.0037) \\
    CER                      & 0.0309$^{**}$  (0.0105)   & -0.0025 (0.0167) \\
    Withdraw                 & 0.2139$^{***}$ (0.0079)  & 0.3219$^{***}$ (0.0126) \\
    Integrations             & -0.1805$^{***}$ (0.0505) & -0.2272$^{***}$ (0.0799) \\
    Mainnet                  & -0.2542$^{.}$ (0.1310)    & -0.3873$^{.}$ (0.2074) \\
    BridgeHack                & -0.0032 (0.0830)          & 0.1386 (0.1314)  \\
    volETH                   & 0.1545 (0.4637)           & -2.0520$^{**}$ (0.7342) \\
    ETH\_APY                 & -0.0122 (0.0198)          & -0.1980$^{***}$ (0.0313) \\
    Stablecoin\_APY          & 0.0518$^{***}$ (0.0029)    & 0.1203$^{***}$ (0.0046) \\
    GasPrice                 & -0.0440 (0.0312)           & -0.0667 (0.0495) \\
    FGI                      & 0.0053$^{***}$ (0.0005)     & 0.0084$^{***}$ (0.0008) \\
    \midrule
    Observations              & 2421              & 2421 \\
    R-squared                & 0.6303             & 0.7023   \\
    \bottomrule
  \end{tabular}
  \raggedright
    Note: This table includes estimates of two panel regression models performed on TVL and TR as dependent variables. The written values represent coefficients while statistical significance levels are shown with stars. Standard errors are reported in parentheses. \\ Significance levels: ˙ p $<$ 0.1, * p $<$ 0.05, ** p $<$ 0.01, *** p $<$ 0.001
\end{table}
 
\begin{table}[htbp]
  \centering
  \caption{Pooled OLS Regression Results}
  \label{tab:pooled_ols_regression_results}
  \begin{tabular}{lcccc}
    \toprule
    & Model 3 (TVL) & Model 4 (TR) \\
    \midrule
    (Intercept)             & 20.0713$^{***}$ (0.2599)   & 4.8802$^{***}$ (0.4700) \\
    BridgeVolume            & 0.0129 (0.0137)           & 0.1057$^{***}$ (0.0247) \\
    Liquidation             & -0.0094$^{***}$ (0.0028)    & 0.0046 (0.0050) \\
    CER                     & 0.1036$^{***}$ (0.0227)  & 0.0932$^{*}$ (0.0410) \\
    Withdraw                & 0.1486$^{***}$ (0.0076)  & 0.2755$^{***}$ (0.0138) \\
    Integrations            & -0.0901$^{*}$ (0.0448) & -0.1066 (0.0810) \\
    Mainnet                 & -0.1378 (0.1158)          & -0.0783 (0.2095) \\
    BridgeHack              & 0.0882 (0.0733)           & 0.1880 (0.1327)  \\
    volETH                  & 0.3214 (0.4124)           & -2.7265$^{***}$ (0.7459) \\
    ETH\_APY                & 0.2545$^{***}$ (0.0340)   & 0.0994 (0.0616) \\
    Stablecoin\_APY         & 0.0619$^{***}$ (0.0031)    & 0.1680$^{***}$ (0.0057) \\
    GasPrice                & -0.0224 (0.0216)          & -0.0366 (0.0391) \\
    FGI                     & 0.0036$^{***}$ (0.0004)     & 0.0048$^{***}$ (0.0008) \\
    \midrule
    Observations            & 807               & 807 \\
    Multiple R-squared      & 0.8154             & 0.8813   \\
    Adj. R-squared          & 0.8126             & 0.8796  \\
    \bottomrule
  \end{tabular}
  \raggedright
  \scriptsize{
    Note: This table reports pooled OLS estimates on the aggregated (cross-network) data. Results are used solely as robustness checks. \\ The written values represent coefficients while statistically significance levels are shown with stars. Standard errors are reported in parentheses. \\ Significance levels: ˙ p $<$ 0.1, * p $<$ 0.05, ** p $<$ 0.01, *** p $<$ 0.001
    }
\end{table}

\begin{table*}[htbp]
  \centering
  \caption{Category-Specific Regressions of TVL (Model 5-6-7) and Revenue (Model 8-9-10) by Network Type}
  \label{tab:network_specific_ols_regression_results_tvl}
  \begin{tabular}{lccccccc}
    \toprule
    & Model 5 (L1) & Model 6 (L2) & Model 7 (AltL1) & Model 8 (L1) & Model 9 (L2) & Model 10 (AltL1) \\
    \midrule
    (Intercept)             & 20.104$^{***}$ (0.256)   & 15.049$^{***}$ (0.301)   & 16.114$^{***}$ (0.405)   & 5.320$^{***}$ (0.445)   & 4.080$^{***}$ (0.435) & 4.866$^{***}$ (0.699)\\
    BridgeVolume            & 0.0469$^{***}$ (0.013)   & 0.170$^{***}$ (0.018)    &  -0.138$^{***}$ (0.021)   & 0.148$^{***}$ (0.022)   & 0.164$^{***}$ (0.025) & -0.197$^{***}$ (0.037)\\
    Liquidation             & -0.007$^{***}$ (0.0017)    & -0.004 (0.004)          & 0.007 (0.005)           & 0.001 (0.003)            & 0.016$^{**}$ (0.006) & 0.036$^{***}$ (0.009)\\
    CER                     & -0.028 (0.039)            & 0.008 (0.009)          & 0.147$^{*}$ (0.060)                & -0.270$^{***}$ (0.069)      & -0.018 (0.013) & 0.475$^{***}$ (0.104)\\
    Withdraw                &  0.116$^{***}$ (0.009)  & 0.118$^{***}$ (0.011)     & 0.393$^{***}$ (0.019)        & 0.216$^{***}$ (0.016)  & 0.142$^{***}$ (0.015) & 0.491$^{***}$ (0.037)\\
    Integrations            & -0.102$^{*}$ (0.044)     & -0.126$^{.}$ (0.068)     & -0.354$^{***}$ (0.100)         & -0.119 (0.077)     & -0.076 (0.099) & -0.443$^{*}$ (0.173)\\
    Mainnet                 & -0.086 (0.115)           & -0.072 (0.177)            & -0.477$^{.}$ (0.260)                    & 0.148 (0.200)           & -0.116 (0.256) & -0.800$^{.}$ (0.449)\\
    BridgeHack              & 0.092 (0.072)            & -0.009 (0.112)           & -0.086 (0.164)                & 0.056 (0.126)            & -0.017 (0.162)  & 0.469$^{.}$ (0.284)\\
    VolETH                  & 0.772$^{.}$ (0.413)    & 0.625 (0.636)           & -0.807 (0.925)              & -1.916$^{**}$ (0.720)    & 1.260 (0.920) & -3.564$^{*}$ (1.596)\\
    ETH\_APY                & 0.257$^{***}$ (0.027)   & 0.609$^{***}$ (0.035)     & -0.104$^{***}$ (0.030)       & 0.304$^{***}$ (0.047)   & 0.721$^{***}$ (0.051) & -0.340$^{***}$ (0.052)\\
    Stablecoin\_APY         & 0.051$^{***}$ (0.003)    & 0.017$^{***}$ (0.004)    & 0.048$^{***}$ (0.005)       & 0.158$^{***}$ (0.005)    & 0.065$^{***}$ (0.006) & 0.096$^{***}$ (0.009)\\
    GasPrice                & -0.012 (0.019)       & -0.095$^{.}$ (0.055)          & -0.098 (0.135)              & 0.022 (0.032)            & -0.150$^{.}$ (0.080)  & -0.174 (0.233)\\
    FGI                     &  0.003$^{***}$ (0.000)   & 0.006$^{***}$ (0.001)    & 0.004$^{***}$ (0.001)         & 0.004$^{***}$ (0.001)     & 0.009$^{***}$ (0.001) & 0.008$^{***}$ (0.002)\\
    \midrule
    Observations            & 807               & 807         & 807                                   & 807             & 807 & 807\\
    R-squared               & 0.7978             & 0.7318    & 0.7208                           & 0.8920             & 0.7513  & 0.7082\\
    Adj. R-squared          & 0.7947             & 0.7278    &  0.7166                            & 0.8904             & 0.7475 &  0.7038\\
    \bottomrule
  \end{tabular}
  \raggedright
  \scriptsize{
    \\ Note: Each column reports estimates from regressions run separately for each network-type using identical specifications. The written values represent coefficients while statistically significance levels are shown with stars. Standard errors are reported in parentheses.
    Significance levels: ˙ p $<$ 0.1, * p $<$ 0.05, ** p $<$ 0.01, *** p $<$ 0.001
  }
\end{table*}

\subsection{Panel Data Results}

Coefficients are rounded to two decimal places due to space constraints. \textbf{Model 1} in \ref{tab:fixed_effects_regression_results} presents \textbf{TVL} regression results for protocols across groups. For L1, we find that a 1\% increase in (\textit{BridgeVolume}) is associated with a significant 0.06\% decrease in TVL for these protocols. In contrast, we find that TVL on L2 has a heterogeneous relationship with bridge activity. For L2, a 1\% increase in bridge volume is associated with a 0.09\% $(-0.06 + 0.15 = 0.09)$ increase in TVL. For the interaction term of AltL1, the relationship is negative and insignificant. The coefficient for \textit{Liquidation} is insignificant, indicating no significant impact of liquidations on TVL. Among control variables, \textit{CER} $(0.03)$, \textit{Withdraw} (0.21), \textit{Stablecoin\_APY} $(0.05)$, and \textit{FGI} $(0.01)$ show statistically significant and positive relationships, while \textit{Integrations} $(0.18)$ show a negative relationship with TVL. \textit{Mainnet} $(-0.254)$ shows a weak significance with a negative relationship. 

\textbf{Model 2} in \ref{tab:fixed_effects_regression_results} presents the regression on \textbf{TR} for protocols in the same network types. A significant coefficient was found for \textit{BridgeVolume}, a 1\% increase in bridge volume increases TR by 0.14\%. For AltL1, we see that TR is negatively related to bridge volume, with a 1\% increase in bridge volume, revenue decreases by 0.13\% $(-0.27 +0.14 = -0.13)$, while \textit{BridgeVolume} has a barely positive relationship with \textit{TR} $(-0.1346 + 0.1381 = 0.0035)$ on L2. The coefficient for \textit{Liquidation} is positive and significant, with a 1\% increase in liquidation, revenue will increase by 0.02\%. Among control variables, \textit{Withdraw} $(0.32)$, \textit{Stablecoin\_APY} $(0.12)$ and \textit{FGI} $(0.01)$ show statistically significant and positive relationships, while \textit{volETH} $(-2.05)$ show a negative relationship on revenue. \textit{ETH\_APY} $(0.19)$ shows statistically significant and negative relationship while \textit{BridgeHack} $(0.14)$ shows no significant relationship with revenue. \textit{Mainnet} $(-0.387)$ shows a weak significance with a negative relationship. 

\subsection{Aggregated and Cross-Category Analysis}

The aggregated OLS regression in \ref{tab:pooled_ols_regression_results} presents a robustness test for \textbf{TVL in Model 3}. The coefficient for \textit{BridgeVolume} is insignificant as expected since cross-chain asset transfers indicate liquidity transfer between chains, not exit for the overall ecosystem. The coefficient of \textit{Liquidation} is significant and negative, indicating that a 1\% increase in liquidations is associated with a 0.01\% decrease in TVL. This result aligns with the traditional view of liquidations as a risk indicator. Regarding the controls, the coefficients for \textit{CER} $(0.10)$, \textit{Withdraw} $(0.15)$, \textit{ETH\_APY} $(0.25)$, \textit{Stablecoin\_APY} $(0.06)$, and \textit{FGI} $(0.004)$ are positive and significant, confirming their role as drivers of TVL. \textit{Integrations} show negative and weakly significant relationship $(0.09)$.

\textbf{Model 4} in \ref{tab:pooled_ols_regression_results} presents robust results for \textbf{TR}. The coefficient for \textit{BridgeVolume} is significant and positive, indicating that a 1\% increase in bridge volume is associated with a 0.11\% increase in TR. The coefficient for \textit{Liquidation} is insignificant. Regarding the controls, the coefficients for \textit{CER} $(0.09)$, \textit{Withdraw} $(0.27)$, \textit{Stablecoin\_APY} $(0.17)$, and \textit{FGI} $(0.01)$ are positive and significant, while \textit{volETH} $(-2.73)$ shows a negative and significant relationship.

Network-type-specific regressions reveal some heterogeneity. The coefficients for \textit{BridgeVolume} in \textbf{TVL (Models 5-6-7)} are significant and positive for L1 $(0.05)$ and L2 $(0.17)$, while AltL1 $(-0.14)$ shows a negative relationship. This reinforces the finding that L2 protocols benefit more from cross-chain activity. The coefficient for \textit{Liquidation} is significant only for L1, indicating that a 1\% increase in liquidations is associated with a 0.01\% decrease in TVL. For controls, \textit{Withdraw}, \textit{Stablecoin\_APY} and \textit{FGI} have a significant and positive relationship for all network categories, while \textit{ETH\_APY} shows significant but heterogeneous coefficients.

\textbf{In Model 8-9-10 for TR}, \textit{BridgeVolume} has a positive and significant relationship in L1 $(0.15)$ and L2 $(0.16)$, while AltL1 $(0.20)$ shows a significant but negative relationship. \textit{Liquidation} has significant and positive relationship in L2 $(0.02)$ and AltL1 $(0.04)$ while it is not significant in L1. \textit{Integrations} show relationship with TR only in AltL1 $(-0.44)$ with a moderate significance. \textit{Withdraw}, \textit{Stablecoin\_APY} and \textit{FGI} are significant and positive drivers for all network types, while \textit{ETH\_APY} has a negative relationship in AltL1 $(0.34)$ but positive in L1 $(0.30)$ and L2 $(0.72)$. \textit{BridgeHack} shows no significance for L2 $(-0.02)$ and weak significance for AltL1 $(0.47)$. \textit{CER} has a significant and positive relationship in AltL1 $(0.48)$ while it has a significant but negative impact in L1 $(-0.27)$.

\section{Discussion}

On-chain finance has evolved into a multi-chain ecosystem, where bridge activity is critical for liquidity flow and protocol growth \cite{Harris2023}. Our results introduce a new dimension to risk management models by incorporating cross-chain activity. We find that the impact of bridge volume on TVL and TR is not uniform and is dependent on the underlying blockchain type.

The positive and statistically significant relationship between bridge volume and TVL on L2 networks might be attributed to the migration of users and capital from L1 to L2 in search of lower fees and higher efficiency. This migration effect mirrors the difference in the number of lending protocol users observed between network types in Figure \ref{fig:users_plot}. The negative relationship between bridge integrations and both TVL and revenue across different network categories suggests that new bridge connections might be acting as liquidity exit channels from lending protocols, rather than growth catalysts for lending protocols. The positive relationship between liquidations and revenue may suggest that protocols have effective fee capture mechanisms during deleveraging events. This aligns with the concept that advanced protocol mechanisms can transform risk events into revenue opportunities.

The behavioral divergence between chains is likely influenced by the composition of the users. L2 networks, with their lower fees, attract a retail-oriented user base that is more sensitive to cross-chain opportunities and market signals, as seen in Figure \ref{fig:users_plot}. In contrast, the Ethereum mainnet, with its entrenched institutional presence, shows greater value in the deposit amount, as seen in Figure \ref{fig:deposit_plot}.

There is a lack of relationship between bridge hacks and dependent variables. Following a hack event, it is surprising to not see that risk-averse users may rapidly deleverage their positions, triggering liquidations and generating fee revenue for protocols in the short term. 

\section{Limitations}

The correlations identified in this analysis are exploratory and do not imply causation. While moderate correlations suggest potential relationships, they do not establish the direction or nature of influence, and the findings should be interpreted within this preliminary framework. This study focuses on the blockchains running the EVM. The observed dynamics may not apply to protocols with different structures or operational models. The insights may not fully generalize to other blockchains with varying transaction costs, user bases, or infrastructure, which can significantly influence protocol behavior. Despite controlling for several key variables (e.g. credit expansion situations, asset price, yields, fear and greed index), the models may suffer from omitted variable bias. This study may include potential endogeneity and simultaneity issues, an instrumental variable method can be used to understand and solve those issues. Data limitations are also a consideration. The analysis relies on available subgraph data, which may include gaps, inconsistencies, or NA values. Variations in data completeness and granularity could affect result comparability across protocols or chains, and the snapshot nature of these data may not capture nuanced financial activities. Additionally, the recent emergence of lending mechanisms means historical data on protocols is limited, potentially constraining the analysis of long-term trends in stability and growth. DeFi space and lending mechanisms are rapidly evolving, this fast-paced development may quickly outdate the findings of this study. This analysis does not include qualitative behavioral insights into user motivations or governance mechanisms, which could have significant impacts on protocol stability and engagement. Such a behavioral analysis would require qualitative data or survey data beyond the scope of this study. Finally, the study focuses on TVL and revenue as primary metrics. While these are standard measures of scale and performance, they do not fully capture protocol risk or resilience. Metrics such as Total Value Redeemable (TVR), additional risk factors such as the severity of bad debt following liquidations and collateral utilization across chains, and attack severity of bridge hacks could offer a more nuanced view of protocol health and stronger insights under cross-chain stress.

\section{Conclusion}

This paper empirically extends existing DeFi stability models by incorporating cross-chain activity variables and analyzing their heterogeneous effects across blockchain network categories. We have investigated the impact of cross-chain bridge activity and liquidation events on TVL and revenue of lending protocols across multiple blockchain categories. Our analysis reveals that cross-chain activity has heterogeneous relationships across different blockchain types. The volume of the bridge has a statistically significant and positive relationship with TVL in L2 networks, while it has a negative relationship with the TVL of Ethereum L1 in the fixed effects regression. The statistical significance disappears in pooled regressions, indicating that fixed effects absorb this crucial heterogeneity. Bridge volume is a significant explanatory variable of revenue across all categories and models, though the direction of this effect varies heterogeneously. We find that an increase in bridge integrations leads to a decrease in TVL and revenue of the lending protocols across blockchain categories that include a total of 9 EVM blockchains. Heterogeneous effects are observed in the bridge volume and liquidations across different blockchain categories in the network-type-specific OLS regressions. All models have moderately high explanatory power, with the R-squared values slightly above the most of the empirical studies in DeFi. The findings are useful for risk assessment frameworks, model and protocol optimizations, guiding protocol cross-chain expansion strategies, and protocol deployment on multiple chains. 

These results demonstrate that cross-chain liquidity flows are a significant factor in the performance and stability of on-chain lending markets. Cross-chain activities and a blockchain-type-aware approach should be taken into account in risk management. The divergent results highlight the importance of a multi-chain strategy and the need for layer-specific risk parameters.

This research could be extended by analyzing directional bridge flows (inflows vs. outflows) to better understand the liquidity source-destination dynamics. Furthermore, an event study could explore the impact of major bridge hacks or failures on interconnected protocol TVL and revenue spillovers.

\label{sec:appendix}

\begin{table}
\centering
\caption{Variance Inflation Factor (VIF) Results}
\label{tab:VIF}
\scriptsize
\begin{tabularx}{\linewidth}{@{}X c X c X c@{}}

\toprule
\textbf{Variable} & \textbf{VIF} & \textbf{Variable} & \textbf{VIF} & \textbf{Variable} & \textbf{VIF} \\
\midrule
BridgeVolume   & 2.08 & Liquidation & 1.26 & Withdraw   & 2.06  \\
GasPrice   & 1.07 &  CER & 1.34 &  ETH\_APY     &  1.19\\
Stablecoin\_APY   & 2.61 & FGI & 1.84 & ETH   & 1.12 \\
\bottomrule
\end{tabularx}

\vspace{0.3em}
\raggedright
\scriptsize
All VIF values are below 5, indicating no multicollinearity concerns.
\end{table}

\begin{table}[t]
\centering
\caption{Context for Dummy Variables}
\label{tab8}

\begin{tabularx}{\linewidth}{l l l}
\toprule
\textbf{Category} & \textbf{Protocol} & \textbf{Date} \\
\midrule

\multirow{5}{*}{Bridge Hacks}
 & ChainSwap   & 2022-11-07 \\
 & Rubic          & 2022-12-25 \\
 & Multichain     & 2023-07-07 \\
 & Orbit Bridge   & 2023-12-31 \\
 & Bungee         & 2024-01-16 \\
\midrule

\multirow{14}{*}{Bridge Integrations}
 & LayerZero      & 2022-10-19 \\
 & Wormhole       & 2022-10-18 \\
 & LayerZero      & 2022-11-30 \\
 & Wormhole       & 2022-11-22 \\
 & Wormhole       & 2022-12-19 \\
 & LayerZero      & 2023-03-10 \\
 & LayerZero      & 2023-03-11 \\
 & LayerZero      & 2023-03-30 \\
 & LayerZero      & 2023-04-26 \\
 & LayerZero      & 2023-04-27 \\
 & LayerZero      & 2023-06-07 \\
 & LayerZero      & 2023-06-14 \\
 & LayerZero      & 2023-07-20 \\
 & LayerZero      & 2023-10-25 \\
\midrule

\multirow{2}{*}{Mainnet Launches}
 & Base        & 2023-08-09 \\
 & zkSync Era  & 2023-03-24 \\
\bottomrule
\end{tabularx}

\vspace{0.3em}
\raggedright
\scriptsize
This table presents the list of bridge hacks, bridge integrations and network launch (mainnet) events, related to the blockchains involved in this study during the research period. Source: Protocol Announcements and DeFiLlama (Last Access: January 1, 2025).
\end{table}

\end{document}